\documentclass[manuscript,natbib]{aastex}

\usepackage[]{natbib}

\newcommand{\vmag}{$V_{mag}$}
\newcommand{\teff}{T$_{\rm eff}$}
\newcommand{\logl}{log(L/L$_{\odot}$)}
\newcommand{\masyr}{mas\,yr$^{-1}$}
\newcommand{\kms}{km\,s$^{-1}$}
\newcommand{\mv}{M$_V$}
\newcommand{\bcv}{BC$_V$}
\newcommand{\msun}{M$_{\odot}$}

\slugcomment{Accepted for publication in Astronomical Journal}
\shorttitle{Alcor}
\shortauthors{Mamajek et al.}
\begin{document}

\title{Discovery of a Faint Companion to Alcor Using MMT/AO 5 $\mu$m Imaging\altaffilmark{1}}

\author{Eric E. Mamajek}
\affil{University of Rochester, Department of Physics \& Astronomy, Rochester, NY, 14627-0171, USA}

\author{Matthew A. Kenworthy, Philip M. Hinz, Michael R. Meyer\altaffilmark{2}}
\affil{Steward Observatory, The University of Arizona, 933 N. Cherry Ave., Tucson, AZ, 85721, USA}


\altaffiltext{1}{Observations reported here were obtained at the MMT
Observatory, a joint facility of the University of Arizona and the
Smithsonian Institution.}
\altaffiltext{2}{Current address: Institute for Astronomy ETH, Physics Department, HIT J 22.4, CH-8093 Zurich, Switzerland}

\begin{abstract}
We report the detection of a faint stellar companion to the famous
nearby A5V star Alcor (80 UMa). The companion has M-band ($\lambda$ =
4.8\,$\mu$m) magnitude 8.8 and projected separation 1''.11 (28 AU)
from Alcor. The companion is most likely a low-mass ($\sim$0.3 \msun)
active star which is responsible for Alcor's X-ray emission detected
by ROSAT (L$_{\rm X}$ $\simeq$ 10$^{28.3}$ erg/s). Alcor is a nuclear
member of the Ursa Major star cluster (UMa; d $\simeq$ 25 pc, age
$\simeq$ 0.5 Gyr), and has been occasionally mentioned as a possible
distant (709'') companion of the stellar quadruple Mizar ($\zeta$
UMa). Comparing the revised Hipparcos proper motion for Alcor with the
mean motion for other UMa nuclear members shows that Alcor has a
peculiar velocity of 1.1 km/s, which is comparable to the predicted
velocity amplitude induced by the newly-discovered companion ($\sim$1
km/s).  Using a precise dynamical parallax for Mizar and the revised
Hipparcos parallax for Alcor, we find that Mizar and Alcor are
physically separated by 0.36\,$\pm$\,0.19 pc (74\, $\pm$\, 39 kAU;
minimum 18 kAU), and their velocity vectors are marginally consistent
($\chi^2$ probability 6\%). Given their close proximity and concordant
motions we suggest that the Mizar quadruple and the Alcor binary be
together considered the 2nd closest stellar sextuplet. The addition of
Mizar-Alcor to the census of stellar multiples with six or more
components effectively doubles the local density of such systems
within the local volume (d $<$ 40 pc).
\end{abstract}

\keywords{
binaries: close --
binaries: general --
binaries: visual --
open clusters and associations: individual (Ursa Major) --
stars: individual (Alcor, Mizar)
}

\facility{MMT, ROSAT, Hipparcos}

\section{Introduction}

Knowing the distribution of companion masses as a function of orbital
separation and primary mass, is fundamental to understanding the
nature of fragmentation of collapsing  molecular cloud cores and star
formation itself.  Evidence suggests that the binary frequency is a
function of stellar mass such that higher mass stars have a higher
binary frequency \citep[c.f.][]{Lada06}.  Whether the distribution of 
companion masses is consistent with having been drawn from the field
star "system" initial mass function across the mass spectrum  of
primaries remains to be demonstrated.  In addition, high order
multiples provides important dynamical constraints to star formation
in clusters and associations \citep{Goodwin07, Parker09}.

The stars Mizar ($\zeta$ UMa) and Alcor (80 UMa) hold an esteemed
place in astronomical lore as perhaps the most famous optical
double. Situated in the middle of the handle of the Big Dipper, Mizar
and Alcor are separated by 11'.8. At this separation, the pair is
resolvable by the naked eye, and indeed the system is famous for its
use in testing vision among many cultures
\citep{Allen1899}. Claims of the physicality of the 
Mizar-Alcor binary varies across the literature, ranging from
confident statements that the two comprise an unphysical ``optical
double'', to the pair being comprised of two unbound members of the
same star cluster (Ursa Major), to being listed as a definite bound
multiple system.

Mizar is resolved in modest telescopes into a 14''.4 binary
\citep{Perryman97} with a probable period of thousands of years. Mizar
A is a nearly equal-mass, double-lined spectroscopic binary with
period 20.54 days and eccentricity of 0.53
\citep{Pourbaix00}. Mizar B is a spectroscopic binary with period 175.57
days and an eccentricity of 0.46 \citep{Gutmann65}. The discovery of
Mizar as a binary is often mistakenly attributed to Giovanni Battista
Riccioli around 1650 \citep[e.g.][]{Allen1899,Burnham78}, however
Galileo's protege and collaborator Benedetto Castelli reported
resolving Mizar in a letter to Galileo dated 7 January 1617.  Galileo
himself resolved the binary and later recorded his measurements on 15
January 1617 \citep{Ondra04, Siebert05}. Besides Alcor, an additional
bright star lies within 8' of Mizar -- the 7th magnitude star HD
116798
\citep[``Stella Ludoviciana'' or ``Sidus Ludovicianum'';][]{Allen1899,
Siebert05}. This star can now be trivially ruled out as being
physically associated with Mizar or Alcor based on its small proper
motion and inconsistent spectrophotometric distance. The ensemble of
Mizar, Alcor, and Sidus Ludovicianum provided the first testing ground
for attempts to solve one of the cosmological conundrums of the 17th
century: trying detecting stellar parallax to confirm the then
controversial heliocentric model. Lodovico Ramponi, in a letter to
Galileo in 1611, sketched out the concept that optical double stars of
different magnitudes (presumed to be identical suns lying at a range
of distances) would provide definite proof of heliocentrism through
the detection of differential parallax.  Galileo sketched an aperture
mask to detect differential parallax in his observations of Mizar,
Alcor, and Sidus Ludovicianum \citep{Siebert05}. Unfortunately for
Galileo, definitive detection of stellar parallax would not be
forthcoming for two more centuries \citep{Bessel1838}.

The Mizar-Alcor system contained further surprises and astronomical
firsts. Mizar A \& B, and Alcor, were together the first resolved
multiple star system photographed on 27 April 1857
\citep{Bond1857}\footnote{150 years to the month before the images
reported in this contribution.}.  While working on the Henry Draper
Memorial project at Harvard College Observatory, Antonia Maury found
Mizar A to be the first spectroscopic binary
\citep[reported by ][]{Pickering1890}. Later, Mizar B was 
reported to be a single-lined spectroscopic binary; independently by
two contemporaneous studies a century ago \citep{Ludendorff08,
Frost08}. Mizar A was also one of the first binary stars to be
resolved using an optical interferometer
\citep{Pease25}.

The Gliese CNS3 catalog lists the Mizar-Alcor system as a ``wide
binary and multiple system'', and the 17th widest multiple system in
their census of solar neighborhood stars \citep{Gliese88}. Alcor is a
bright (\vmag\, = 3.99) A5Vn star situated 708''.55 from Mizar A
\citep{Gray89, Fabricius02}. Alcor's properties are summarized in
Table
\ref{tab:alcor}. Alcor was also detected as an X-ray source in both
pointed observations and the All-Sky Survey of the ROSAT X-ray
observatory \citep{Voges99, ROSAT00, White00}. X-ray emission is rare
among A-type stars \citep{Simon95,Schroder07}. X-ray emission among
A-type stars is often proposed to be emitted from the coronae of
low-mass companions, and indeed the majority of X-ray emitting A-type
stars show some signs of multiplicity \citep{Schroder07}.

Alcor has been reported to have rapid
radial velocity variations \citep{Frost08, Heard49} -- possibly
intrinsic to the star itself. The star is often flagged as ``SB''
\citep{Johnson53, Hoffleit64, Gliese91}, but no orbit has ever been
reported.
\citet{Frost08} stated that there is ``no doubt that Alcor
is also a spectroscopic binary'', and that the displacement and
multiplicity of the Mg $\lambda$4481 and Balmer lines ``succeed each
other so rapidly that I have found it necessary to have spectrograms
of this star made in continuous succession for several hours''. Frost
did not publish his radial velocities. \citet{Heard49} stated ``there
is a fair degree of probability that Alcor varies in velocity''. Their
observations over a $\sim$9 year span appear to be of very low quality
(two observations 4 minutes apart had radial velocities that differed
by 33 \kms). \citet{Heard49} estimated a velocity amplitude of 6
\kms\, (no uncertainty), but no period was reported. 
The most recent comprehensive assessment of the binarity of Alcor was
reported by \citet{Abt65}. In his large survey of A-star multiplicity,
\citet{Abt65} measured 13 additional radial velocities in 1959-1961,
and said the radial velocities ``show a slightly excessive scatter.''
Taking into account previously published velocities, Abt concluded
that Alcor's velocity was ``Constant:'', and he considered the star to
be single.

\begin{deluxetable}{lcl}
\setlength{\tabcolsep}{0.03in}
\tablewidth{0pt}
\tablecaption{Properties of Alcor \label{tab:alcor}}
\label{tab:alcor}
\tablehead{Property & Value & Ref.}
\startdata 
Parallax       & 39.91\,$\pm$\,0.13 mas      & 1\\
Distance       & 25.06\,$\pm$\,0.08 pc       & 1\\
$\mu_{\alpha}$ & 120.21\,$\pm$\,0.12 \masyr  & 1\\
$\mu_{\delta}$ & -16.04\,$\pm$\,0.14 \masyr  & 1\\
RV             & -9.6\,$\pm$\,1.0 \kms       & 2\\
V$_{\rm mag}$  & 3.99 mag                    & 1\\
B-V            & 0.169\,$\pm$\,0.006 mag     & 1\\
V-I$_c$        & 0.19\,$\pm$\,0.03 mag       & 1\\
L'$_{\rm mag}$ & 3.65 mag                    & 3\\
\teff\,        & 8030 K                      & 4\\
Spec. Type     & A5Vn                        & 5\\
BC$_V$         & -0.02 mag                   & 6\\
M$_V$          & 2.00\,$\pm$\,0.01 mag       & 7\\
\logl\,        & 1.11\,$\pm$\,0.01 dex       & 8\\
L$_X$          & 10$^{28.34}$ erg\,s$^{-1}$  & 9\\
Mass           & 1.8 \msun                   & 10\\
Age            & 0.5\,$\pm$\,0.1 Gyr         & 11\\
$U, V, W$      & +14.3, +2.7, -9.3 \kms      & 12\\
               & ($\pm$0.5, 0.7, 0.6 \kms)   & 12\\ 
\enddata

\tablecomments{References: (1) \citet{vanLeeuwen07}, 
distance is inverse of parallax,
(2) \citet{Gontcharov06}, (3) \citet{Kidger03}, (4) mean
value from \citet{Blackwell98}, \citet{Cenarro01}, 
\citet{Gray03}, and \citet{LeBorgne03}, (5) 
\citet{Gray89}, (6) from adopted \teff\, and tables of \citet{Flower96}, 
(7) from
adopted V magnitude and parallax, assuming zero extinction, (8) from
adopted
\mv\, and \bcv, (9) soft X-ray luminosity (0.2-2.4 keV) in ROSAT band,
calculated using count-rate and hardness ratio HR1 from
\citet{Voges00}, energy conversion factor from \citet{Fleming95},
and the adopted parallax, (10) combining \teff\, and \logl\, values
with z=0.02 evolutionary tracks of \citet{Lejeune01}, (11) UMa 
cluster age from \citet{King03}, (12) Galactic Cartesian velocity
vector, calculated in \S3.4.1.}
\end{deluxetable}

In this paper, we (1) report the discovery of a faint companion to
Alcor at separation 1''.1 with the 6.5-m MMT telescope using the
adaptive secondary (MMT/AO), (2) argue that the nature of the
companion is most likely a low-mass dwarf which is also responsible
for Alcor's X-ray emission detected by ROSAT and subtle peculiar
motion with respect to the mean motion for Ursa Major nucleus members,
and (3) present evidence that the astrometry of the Mizar-Alcor system
is consistent with the Mizar quadruple and Alcor double being
physically associated, making the Mizar-Alcor a probable sextuplet,
and the 2nd closest such multiple known.

\section{Observations \label{sec:obs}}

As part of a recently completed survey to image brown dwarf and
exoplanet companions to nearby intermediate-mass stars \citep[Mamajek et
al., in prep.; ][]{Kenworthy09}, the star Alcor was imaged with
the Clio 3-5\,$\mu$m imager in conjunction with the adaptive secondary
mirror on the 6.5-m MMT telescope \citep{Brusa04}. Clio is a high
well depth InSB detector with 320 $\times$ 256 pixels and 49 mas
pixels and field of view of 15''.6 $\times$ 12''.4 at M-band when
attached to the MMT \citep{Sivanandam06, Hinz06, Heinze08}.

Alcor was imaged at M-band with Clio and MMT/AO on 08 Apr 2007 (start
UT 08:26) for a total integration time of 2697 sec (0.75
hr). Observations were stopped due to cloud cover.  Alcor was
beam-switch nodded 5''.5 along the long axis of Clio after each 5
images. The observations of Alcor consist of a series of 129 images of
20.91 second exposures; each consisting of 100 coadded frames of 209.1
msec. The short exposure time was selected to keep the sky background
counts below the nonlinearity threshold for Clio ($\sim$40k ADU). The
primary star is unsaturated in all the frames, with a peak count value
of approximately 3800 counts above the local background level, and a
PSF with a full width half maximum of 4.0 pixels (0''.20). The Clio
images were taken with a Barr Associates M-band filter with half power
range of 4.47-5.06 $\mu$m and central peak wavelength of 4.77\,$\mu$m.

\begin{deluxetable}{ll}
\setlength{\tabcolsep}{0.03in}
\tablewidth{0pt}
\tablecaption{Photometry and Astrometry for Alcor B \label{tab:alcorb}}
\tablehead{Property & Value}
\startdata    
$\Delta$M     & 5.175 $\pm$ 0.013 mag \\
m$_{M}$       & 8.82 $\pm$ 0.05 mag \\
M$_{M}$       & 6.83 $\pm$ 0.05 mag \\
PA ($\theta$) & 208$^{\circ}$.82 $\pm$ 0$^{\circ}$.08 \\
sep ($\rho$)  & 1109.5 $\pm$ 2 mas (27.8 AU @ 25.1 pc)\\
epoch         & JD 2454199.35 (J2007.267) 
\enddata
\end{deluxetable}

\section{Analysis \label{sec:anal}}

\subsection{Astrometry}

We use a custom pipeline to reduce Clio data, with steps including
automatic amplifier noise pattern correction and beamswitching
\citep[described in ][]{Kenworthy09}. Bad pixels are interpolated over
with a 3$\times$3 pixel median filter. The science images are
resampled with bilinear interpolation, and rotated with North at the
top of the image and East to the left.

We use observations of the triple system HD 100831 (HIP 56622; STF
1553AB) to calibrate the plate scale and orientation of the
detector. The system consists of a single primary star and a
spectroscopic, unresolved (separation $<$ 1 mas) binary system with a
period of approximately 3000 years. The primary and secondary are
separated by approximately 6.1 arcseconds.  This system has been
observed over several epochs ranging back to 1890, showing that the
orbital motion is closely approximated by a linear trend in position
angle and angular separation. We use astrometry from Hipparcos
\citep{Perryman97} and from \citet{Sinachopoulos07} from 1990
through to 2005 to extrapolate the PA and separation at the
observation epoch. We predict that the position angle of the HD 100831
binary at epoch 2007.267 was 165$^{\circ}$.74 $\pm$ 0$^{\circ}$.08
with separation 6''.136 $\pm$ 0''.010.  Using these values we
calculate the plate scale and orientation of the Clio detector on the
April 2007 run, two days after carrying out the Alcor B observations.
Our plate scale (48.56 $\pm$ 0.10 mas\, pix$^{-1}$) is similar to the
plate scales determined during other Clio observation runs. The
Position Angle offset for Clio differs from previous runs by 0.5
degrees, consistent with the repeatability of mounting Clio over
several runs. The errors in the measurement of Alcor B astrometry is
dominated by the astrometric uncertainty in the orbit of HD 100831.

Alcor B is clearly seen in all 129 science images. We determine the
position offset and magnitude difference between A and B by using
Alcor A as a reference PSF for each of the frames. Alcor B sits in the
halo of uncorrected light from Alcor A, and so we estimate the local
background about Alcor B by removing the azimuthal median of a set of
nested concentric rings centered on Alcor A out to a radius of
3''. The reference PSF is then scaled in intensity and translated over
to the location of Alcor B, and subtracted off. We then use a custom
fitting routine to explore this three parameter space (X and Y
offsets, plus the magnitude difference) by minimizing the residuals of
this subtraction in a circular aperture centered on the position of
Alcor B, using Alcor A as a PSF reference. Since we are able to use
the unsaturated image of Alcor A as our PSF reference, we do not have
to approximate the PSF of Alcor B or make any other simplifying
assumptions, so we use an iterative process to determine the best fit
parameters. If the fitting routine does not converge to a solution
within 40 iterations, the fit is discarded (79 images are
retained). Including the astrometric uncertainties determined from the
calibrator binary, the mean values are separation $\rho$ = 1''.1095
$\pm$ 0''.0020 and position angle $\theta = 208^{\circ}.82
\pm 0^{\circ}.08$. 

\subsection{Photometry}

Using the same fitting routine, we measure a magnitude difference of
$\Delta M=5.175 \pm 0.013$ mag with respect to Alcor A.  The absolute
photometric uncertainty for Alcor B is dominated by the uncertainty in
the M-band magnitude for Alcor A, which is unmeasured. Alcor A is an
A5Vn star with negligible reddening. Combining its L' magnitude
\citep[3.65; ][ we assume $\pm$0.01 mag uncertainty]{Kidger03} with
the predicted intrinsic L'-M color for A5V stars \citep[0.01;
][]{Bessell88}, and assuming a conservative total uncertainty in the
intrinsic color and photometric conversion of $\pm$0.05 mag, we
estimate the M magnitude of Alcor to be 3.64 $\pm$ 0.05 mag. This
leads to an apparent M-band magnitude for Alcor B of m$_M$ = 8.82
$\pm$ 0.05 mag.

\begin{figure}
\epsscale{.80}
\includegraphics[height=3in,angle=-90]{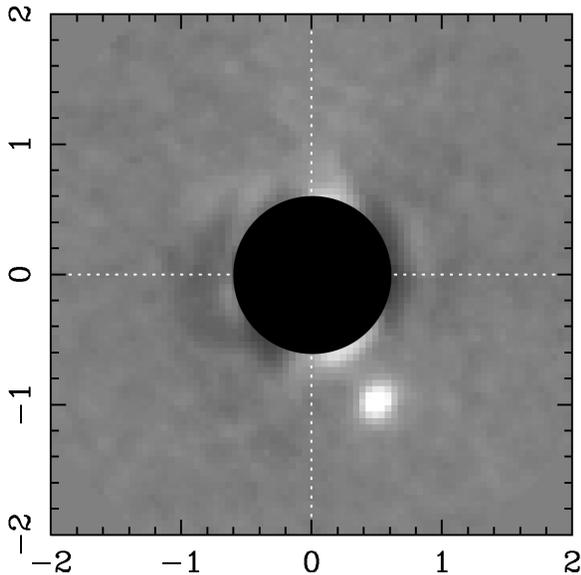}
\caption{Image of Alcor B as imaged with Clio at the MMT. The ordinate
and abscissa are RA and Dec offset from Alcor, respectively, in
arcseconds. North is up and east is left, and the inner 0''.6 radius
of the subtracted PSF of Alcor A is masked. Ten exposures of 20.1
seconds are averaged together to form a total exposure of 201
seconds. Background subtraction is carried out by removing the
azimuthal medians of annuli centered on Alcor A. The color scale is
from -15 to +15 counts. 
\label{fig:alcorb}}
\end{figure}

\subsection{X-ray Emission}

Alcor has an X-ray counterpart in the ROSAT All Sky Survey \citep[1RXS
J132513.8+545920; ][]{Voges00} situated 4'' away from Alcor's optical
position, but with X-ray positional uncertainty of 13''. The total
exposure time was 552 seconds, and the RASS observations were taken
between 27 Nov 1990 and 1 Dec 1990. Using the soft X-ray counts in the
ROSAT band (0.2-2.4 keV) and hardness ratio HR1 from \citet{Voges00},
and using the energy conversion factor from
\citet{Fleming95}, and the adopted parallax from \citet{vanLeeuwen07},
we estimate an X-ray luminosity of 10$^{28.28}$ erg\,s$^{-1}$. The
hardness ratio HR1 is defined following \citet{Schmitt95} and
\citet{Voges99} as HR1 = (B-A)/(B+A), where A is the ROSAT X-ray
count-rate in the 0.1-0.4 keV band, and B is the count-rate in the
0.5-2.0 keV band. Alcor was also detected by ROSAT in a 2608 sec
observation on 8 May 1992, and is reported in the Second ROSAT Source
Catalog of Pointed Observations \citep{ROSAT00} as X-ray source 2RXP
J1325.9+545914. No position error is given, but the X-ray source is
3''.5 away from Alcor, and given typical ROSAT positional
uncertainties, it is extremely likely that the Alcor system is
responsible for the X-ray emission. Using the soft X-ray counts in the
ROSAT band (0.04679 ct s$^{-1}$; 0.2-2.4 keV) and the reported
hardness ratio (HR1 = -0.37), and using the energy conversion factor
from \citet{Fleming95}, and the adopted parallax from
\citet{vanLeeuwen07}, we estimate an X-ray luminosity of 10$^{28.35}$
erg\,s$^{-1}$. 

We adopt an exposure time-weighted mean ROSAT X-ray luminosity of
10$^{28.34}$ erg\,s$^{-1}$. Independently, and using the same archival
ROSAT data, \citet{Schroder07} report Alcor as an unresolved ROSAT
X-ray source with luminosity L$_X$ = 10$^{28.27}$ erg\,s$^{-1}$. This
is only 17\%\, lower than the mean value we calculate, but within the
systematic uncertainties for X-ray luminosity estimation using ROSAT
count rates and hardness ratios.

\subsection{Kinematic Information}

\subsubsection{Velocity of Alcor}

Combining the position, proper motion, and parallax from the revised
Hipparcos \citep{vanLeeuwen07} with the radial velocity from the
compiled catalog of \citet{Gontcharov06}, we estimate the velocity of
Alcor in Galactic Cartesian coordinates to be $U, V, W$ = +14.2, +3.0,
-9.4 \kms ($\pm$0.4, 0.7, 0.6 \kms). The best modern long-baseline
proper motion for Alcor comes from the Tycho-2 catalog
\citep{Hog00}, and combining the revised Hipparcos parallax and 
\citet{Gontcharov06} radial velocity with the Tycho-2 proper 
motion gives a velocity of $U, V, W$ = +14.3, +2.7, -9.3 \kms
($\pm$0.5, 0.7, 0.6 \kms), i.e. negligibly different ($<$0.3 \kms\,
per component) from that calculated using the short-baseline revised
Hipparcos proper motion.

\subsubsection{Velocity of Mizar}

In order to calculate an accurate center-of-mass velocity for the
Mizar quadruple, we need an estimate of the systemic radial velocity
for the system. The mass of Mizar B and its companion is not
well-constrained, so it is difficult to calculate an accurate systemic
velocity for Mizar.  The systemic velocity of Mizar A is
-6.3\,$\pm$\,0.4 \kms\,
\citep{Pourbaix00} and that for B is -9.3\,$\pm$\,0.1 \kms\, \citep{Gutmann65}.
Guttman (1965) estimates that the Mizar B binary is $\sim$80\% of the
mass of the Mizar A binary. Adopting the mass of the Mizar A binary
(4.9 \msun) from \citet{Hummel98}, then the mass of Mizar B is likely
to be $\sim$3.9 \msun. Using these masses, we can estimate a
mass-weighted systemic radial velocity of the Mizar AB quadruple
system of -7.6 \kms, with a conservative uncertainty of $\sim$1
\kms.

We combine the revised Hipparcos trigonometric parallax from
\citep[38.01\, $\pm$\, 1.71 mas][]{vanLeeuwen07} and the dynamical
parallax from \citep[39.4\, $\pm$\, 0.3 mas][]{Hummel98} to estimate a
weighted mean parallax of $\varpi$ = 39.36\, $\pm$\, 0.30 mas.
Using this systemic radial velocity, the weighted mean parallax, and
the proper motion from \citet{vanLeeuwen07}, we calculate a velocity
of Mizar of $U$, $V$, $W$ = 14.6, 3.1, -7.1 \kms\, ($\pm$0.5, 0.7, 0.6
\kms).

\subsubsection{Velocity of Ursa Major Star Cluster}

\begin{figure}
\epsscale{.80}
\includegraphics[height=3in,angle=0]{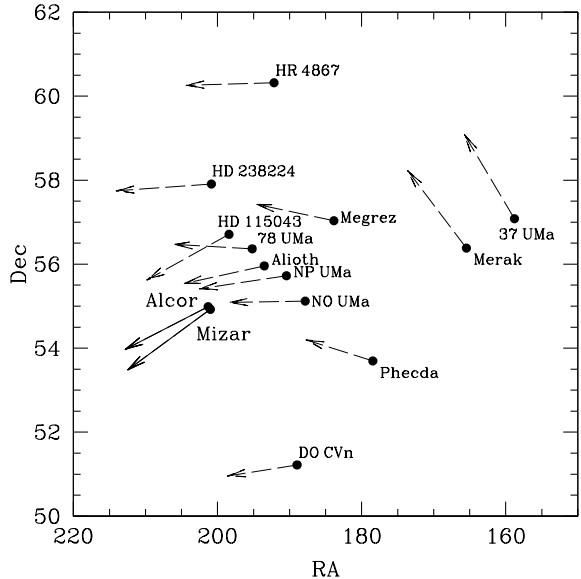}
\caption{Map of the UMa nuclear members in equatorial coordinates.
Arrows indicate 200 kyr of proper motion. Census of UMa nucleus
members comes from \citet{King03}, except for HD 238224 (Mamajek, in
prep.). The dispersion amongst the (mostly convergent) proper motion
vectors comes from a mix of geometric projection effects (the stars
are at high declination) and peculiar motions (dispersion in
tangential motions at the $\sim$1 \kms\, level). Mizar, Alcor, and the
central, most massive UMa nucleus member Alioth have distances of
25.4\,$\pm$\,0.2 pc, 25.1\,$\pm$\,0.1 pc, and 25.3\,$\pm$\,0.1 pc,
respectively. Using the revised Hipparcos parallaxes
\citep{vanLeeuwen07}, the mean distance to the 14 UMa nucleus systems
is 25.2\,$\pm$\,0.3 pc -- implying that Mizar-Alcor is codistant with
the other nuclear members.
\label{fig:map}}
\end{figure}

From the revised Hipparcos astrometry \citep{vanLeeuwen07}, published
mean radial velocities \citep{Gontcharov06}, and nucleus membership
from \citet{King03}, we find the mean velocity vector of the UMa
nucleus to be $U$, $V$, $W$ = 15.0, 2.8, -8.1 ($\pm$ 0.4, 0.7, 1.0)
\kms, a convergent point of $\alpha$, $\delta$ = 300$^{\circ}$.9,
-31$^{\circ}$.0 with S$_{tot}$ = 17.3 $\pm$ 0.6 \kms. Our UMa cluster
velocity compares well to the unweighted mean measured by
\citet{King03}: $U$, $V$, $W$ = 14.2, 2.8, -8.7 ($\pm$ 0.7, 1.3, 1.8)
\kms.  A figure showing the positions and proper motion vectors for
the UMa nuclear members is shown in Figure
\ref{fig:map}. 

Using the calculated velocity vectors for Alcor, Mizar, and the UMa
cluster, we find that Alcor shares the motion of UMa to within 1.4 $\pm$
1.6 \kms, and Mizar shares the motion of UMa to within 1.3 $\pm$ 1.7
\kms. Hence both Alcor and Mizar are consistent with being kinematic
UMa members (although we discuss the intrinsic velocity dispersion of
the group further in \S\ref{pec}). Subtracting the motion of Alcor
from that of Mizar yields $\Delta U$, $\Delta V$, $\Delta W$ = -0.4,
0.0, -2.4 \kms\, ($\pm$0.7, 1.0, 0.9
\kms), and a difference in motion of 2.7\,$\pm$\,0.8 \kms. 
Testing the hypothesis that the motion of Alcor is consistent with
that of Mizar, the difference results in $\chi^2$/d.o.f. = 7.4/3 and a
$\chi^2$ probability of 6\%. Hence, the motion of Alcor and Mizar are
consistent at the $\sim$2$\sigma$ level, given the observational
uncertainties.

We find both Alcor and Mizar to be comoving within 1.5 \kms\, of the
mean UMa cluster motion. {\it What is the probability that a field
A-type star would have a velocity as similar as Alcor's and Mizar's is
to the UMa nucleus?} To answer this question, we
cross-referenced the revised Hipparcos astrometry catalog
\citep{vanLeeuwen07} with the \citet{Gontcharov06} compiled radial
velocity catalog, and calculate UVW velocities for A-type stars
\citep[spectral types from ][]{Perryman97} with parallaxes of $>$10
mas (d $<$ 100 pc) and parallax uncertainties of $<$12.5\%. Given
these constraints, we compile a catalog of velocities for 1018 A-type
stars, 6 of which are known UMa nucleus members. After removing the 6
UMa A-type nucleus members, we find that only 1 A-type star within 100
pc (HIP 75678) has a velocity within 2 \kms\, of the UMa nucleus
(1/1012 $\simeq$ 0.1\%). The typical error in the space motions for
the A-type field stars is $\sim$2.5 \kms ($\sim$1.4 \kms\, per
component). We find that only 2.5\% (25/1012) of field A-type stars
have motions within 5 \kms\, of the UMa velocity vector. Hence, given
its velocity alone, a conservative upper limit to the probability that
Alcor might be an interloper to the UMa cluster is probably in the
range of $\sim$0.1-2.5 \%.

\subsubsection{Peculiar Motion of Alcor \label{pec}}

Independent of the radial velocity values, we can test how
consistent Alcor's tangential (proper) motion is with UMa membership. 
Following the techniques discussed in
\citet{Mamajek05}, we find that Alcor's revised Hipparcos proper
motion toward the UMa convergent point is $\mu_{\upsilon}$ = 120.9
$\pm$ 0.1 \masyr, and the perpendicular motion is $\mu_{\tau}$ = 9.2
$\pm$ 0.1 \masyr. At Alcor's distance this translates into a peculiar
motion of 1.1 $\pm$ 0.1 \kms.

{\it What velocity dispersion do we expect among the UMa nuclear
members if the cluster is in virial equilibrium?} From the census of
UMa nuclear members from \citet{King03} and the astrometry from
\citet{vanLeeuwen07}, we estimate that the stellar mass of the UMa
nucleus is approximately $\sim$28 M$_{\odot}$ and encloses a volume of
$\sim$100 pc$^{3}$. The predicted 1D virial velocity for this stellar
system is 0.1 \kms, suggesting that the peculiar velocity of 1.1 $\pm$
0.1 \kms\, is significantly deviant. However, the distribution of
peculiar motions for the rest of the UMa nucleus members, using the
revised Hipparcos proper motions and the mentioned convergent point,
is consistent with a 1D velocity dispersion of 1.1 $\pm$ 0.2 \kms,
implying that Alcor's peculiar motion is not unusual compared to the
other nuclear members. Our estimate of the 1D velocity dispersion is
within the errors of that estimated by \citet{Chupina01} of 1.33
\kms\, (no uncertainty). The UMa nucleus has 9 stars with peculiar
velocities of $<$0.5 \kms, while the other outliers have peculiar
velocities between 0.9 and 4.4 \kms. {\it All} of these UMa nucleus
stars with peculiar motions of $>$0.5 \kms\, have been claimed to be
stellar multiples (including HD 109011, 111456, 113139, 238224, Mizar,
and now Alcor). So the likely reason that the observed 1D velocity
dispersion of the UMa nucleus is $\sim$10$\times$ the predicted virial
velocity is probably due to the effects of stellar multiplicity on the
proper motions, rather than this long-lived $\sim$0.5 Gyr-old nucleus
being unbound.

Given Alcor's position in the UMa nucleus, similarity of motion with
other UMa nucleus members, proximity to UMa member Mizar, HR diagram
position consistent with other UMa members, and the inherent low space
density of A dwarfs (local density is $\sim$10$^{-3}$
pc$^{-3}$)\footnote{Calculated using the census of stars within 10 pc
from the Henry et al. RECONS project:
http://www.chara.gsu.edu/RECONS/.}, it is extremely unlikely that
Alcor could be an interloper. We conclude that Alcor is an UMa member,
and that its motion is plausibly perturbed at the 1 \kms\, level by
the newly discovered companion star.

\section{Discussion}

\subsection{The Nature of Alcor B}

We investigate three scenarios for the nature of Alcor B:
(1) interloper, (2) white dwarf bound companion, (3) low-mass main
sequence bound companion. If the companion is bound, its age should be
identical to that of Alcor A \citep[i.e. 0.5 $\pm$ 0.1 Gyr;][]{King03},
and its apparent magnitude translates into an absolute M-band
magnitude of M$_M$ = 6.83 $\pm$ 0.05 (adopting Alcor's parallax of
$\varpi$ = 39.91\,$\pm$\,0.13 mas).

$\bullet$ Scenario 1 (interloper): The companion is very bright for a
background object. Alcor is at high Galactic latitude ($b$ =
+61$^{\circ}$.5). The number of M-band (approximately the same as IRAC
4.5 $\mu$m) background stars can be estimated from Fig. 1 of
\citet{Fazio04}. An approximate fit to the differential number counts
in the Bootes field ($b$ = +67.3) due to stars is log$_{10}$(dN/dM)
[num mag$^{-1}$ deg$^{-2}$] $\simeq$ -2.0 + 0.33\,mag$_{4.5\,\mu
m}$. The predicted density of background stars brighter than mag$_M$
$<$ 8.8 is $\sim$12 deg$^{-2}$, and the number predicted within 1''.11
of Alcor is $\sim$3$\times$10$^{-6}$. In our initial imaging survey of
$\sim$20 such A-type stars, we would have expected to find
$\sim$6$\times$10$^{-5}$ interlopers of brighter magnitude and closer
proximity. We also empirically measure the density of K$_s$-band
($\lambda$ = 2.2\,$\mu$m) stars brighter than K$_s$ mag of 8.8 near
Alcor in the 2MASS catalog \citep{Cutri03}, and find 10
deg$^{-2}$. Since most stars have K$_s$-M colors of $\sim$0.0, the
2MASS K$_s$ density provides a useful check on the differential number
counts provided by \citet{Fazio04}. If the star is a background star,
it does not provide an explanation for Alcor's X-ray emission or
peculiar motion with respect to the UMa nucleus. We ascribe a
negligible probability ($\sim$10$^{-4.2}$) that Alcor's faint
companion is a background star.

$\bullet$ Scenario 2 (white dwarf): Given the age of the UMa cluster,
any members whose initial mass was originally $\sim$2.9-7 M$_{\odot}$
are now white dwarfs, most likely in the mass range $\sim$0.7-1.1
M$_{\odot}$ \citep{Lejeune01,Kalirai09}. If we
hypothesize that Alcor B was originally a 0.5 Gyr-old 2.9 M$_{\sun}$ star, it
should now be a cooling 0.7 M$_{\odot}$ white dwarf star
\citep{Kalirai09}.  The white dwarf cooling tracks of 
\citet{Bergeron95}\footnote{http://www.astro.umontreal.ca/$\sim$bergeron/CoolingModels/}
do not include M-band, but does include K-band. If we assume K-M color
of zero, then $M$ $\simeq$ $K$ $\simeq$ 8.8 implies a white dwarf
cooling age of $\sim$270 kyr and a predicted T$_{\rm eff}$ $\simeq$
100,000 K.  While we can not completely rule out the companion being a
white dwarf with the data in hand, we can estimate a rough probability
for B being a white dwarf: P $\sim$ N$_{stars}$
$\Delta\tau_{WD}$/$\tau_{age}$ $\sim$ 0.01, where N$_{stars}$ is the
number of stars in the UMa nucleus ($\sim$20), $\Delta\tau_{WD}$ is
the time interval of rapid evolution that we are concerned with (the
white dwarf cooling timescale), and $\tau_{age}$ is the age of the
cluster \citep[0.5 $\pm$ 0.1 Gyr;][]{King03}.  While a white dwarf
companion might explain Alcor's peculiar motion, it does not explain
the X-ray emission, and it appears very unlikely (P $\sim$ 10$^{-2}$)
that we would serendipitously discover a very luminous, hot, white
dwarf companion during this very short period of its evolution.

$\bullet$ Scenario 3 (low-mass dwarf companion): Using the log(age/yr)
= 8.7 evolutionary tracks of \citet{Baraffe98}, a low-mass star with
absolute M magnitude of 6.83 translates into a mass of 0.30
M$_{\odot}$ (and predicted \teff\, = 3437\,K, \logl\, = -1.99, L$_{\rm
bol}$ = 10$^{31.60}$ erg\,s$^{-1}$, spectral type $\sim$M2V). If the
low-mass dwarf is responsible for the ROSAT X-ray emission (L$_X$ =
10$^{28.34}$ erg\,s$^{-1}$), then log(L$_X$/L$_{bol}$) = -3.26. Such
an X-ray luminosity is typical of M dwarfs members of the similarly
aged (625 Myr) Hyades cluster \citep{Stern95}. {\it The ROSAT X-ray
emission of Alcor may be parsimoniously explained by the existence of
a low-mass active companion.} If the observed orbital separation
corresponds to the semi-major axis (27.8 AU), then A (with mass 1.8
\msun) and B (with mass 0.3 \msun) would have velocity amplitudes of
1.2 \kms\, and 7.0
\kms, respectively, and a predicted period of $\sim$100 yr. 
Remarkably, the predicted velocity amplitude for Alcor A is similar in
magnitude to the measured peculiar motion of Alcor A with respect to
the Ursa Major nucleus mean motion. It is doubtful that
the observed companion could be responsible for the unconfirmed
radial velocity variations observed over a 9-yr period by \citet{Heard49}.

The hypotheses that the new companion is a background star or a white
dwarf companion appears to be very low, with approximate probabilities
of $\sim$10$^{-4}$ and $\sim$10$^{-2}$, respectively. Not only is the
idea of the companion being physical very likely, but it provides a
likely explanation for why Alcor is an X-ray source at the observed
X-ray luminosity, and why Alcor's velocity is peculiar with respect to
the Ursa Major mean motion at the $\sim$1 \kms\, level.  We conclude
that the companion is likely to be physical, and a low-mass ($\sim$0.3
\msun) dwarf.

\subsection{Mizar-Alcor: A Hierarchical Sextuplet}

While Mizar and Alcor was considered a wide-separation binary by
\citet{Gliese88}, the two stars were claimed to belong to
different kinematic subunits within the UMa cluster by
\citet{Chupina01}.  As the question of whether Mizar and Alcor
comprise a physical binary appears to be unanswered, we decided to
explore the issue using modern astrometric data. We do this by
exploring the extent to which Mizar and Alcor are comoving and
codistant, and testing whether they could be a bound system.

{\it How likely is it that two UMa nucleus members (e.g. Mizar and
Alcor) would lie within 709'' of each other but not constitute a
multiple system}? The UMa nucleus contains 15 systems within a
$\sim$200 deg$^{2}$ region of sky (density of $\sim$0.8 stars
deg$^{-2}$). Hence the number of predicted UMa members within 709'' of
a random UMa member is $\sim$0.1. So Mizar and Alcor are projected
unusually close to one another if they do not constitute a physical
subsystem, but are both UMa members.

{\it To what degree are Mizar and Alcor consistent with being
co-distant?} For calculating distances to Alcor and Mizar, 
we adopt the parallax for Alcor listed in Table 1
\citep[39.91\, $\pm$\, 0.13 mas;][]{vanLeeuwen07} and the
parallax for Mizar calculated in Sec. 3.4.2 (39.36\, $\pm$\, 0.30
mas).  The parallaxes are consistent with distances of
25.4\,$\pm$\,0.2 pc for Mizar and 25.1\,$\pm$\,0.1 pc for Alcor,
respectively, and only differ by 2.7$\sigma$. Monte Carlo modeling of
the parallax uncertainties leads to a physical separation between the
Mizar and Alcor systems of $\Delta$ = 0.36\,$\pm$\,0.19 pc (74\,
$\pm$\, 39 kAU). The minimum possible separation is $\Delta_{min}$ =
17.8 kAU. For reference, the most massive and central UMa member --
Alioth -- lies at $d$ = 25.3 $\pm$ 0.1 pc \citep[from revised
Hipparcos parallax;][]{vanLeeuwen07}, and so is statistically
consistent with being codistant with Mizar-Alcor.  Using the adopted
distances, Alcor is physically 2.01 $\pm$ 0.02 pc away from the
central UMa star Alioth.

We already demonstrated in \S3.4.3 that Alcor and Mizar differ in
motion by only 2.7$\pm$0.8 \kms, and are marginally statistically
consistent with co-motion.  {\it What orbital velocities would we
expect for the Alcor binary and Mizar quadruple?} If we assume a total
mass for the Mizar system of $\sim$9 \msun, a total mass of $\sim$2
\msun\, for the Alcor binary, and a presumed orbital semimajor axis of
74 kAU, then one would predict relative orbital velocities of
$\sim$0.3 \kms\, for the Alcor center-of-mass and $\sim$0.07 \kms\,
for the center-of-mass of the Mizar quadruple.  If Alcor and Mizar are
actually at their minimum possible separation (17.8 kAU), then the
velocity amplitudes would be $\sim$0.6 \kms\, (Mizar) and $\sim$0.1
\kms\, (Alcor). Hence the center-of-mass motions of Alcor and Mizar
are likely to be within $<$0.7 \kms\, along any axis, and within the
uncertainties of the current astrometric measurements.

\section{Summary}

We conclude that a low-mass main sequence companion physically bound
to Alcor A is the most likely explanation for the nature of Alcor
B. Future observations confirming common proper motion, and multiband
imaging or spectroscopy confirming that the companion is indeed a
M-type dwarf, are necessary to confirm this hypothesis. The newly
discovered companion is unlikely to be responsible for the short
timespan radial velocity variations observed by \citet{Frost08} and
\citet{Heard49}. The case for the Alcor binary and the Mizar 
quadruple constituting a bound sextuplet with physical separation
$\Delta$ = 0.36\,$\pm$\,0.19 pc (74\, $\pm$\, 39 kAU) is also strong,
given the statistical consistency of their space velocities.

Recent simulations of multiple star evolution in dense stellar
clusters by \citet{Parker09} shows that clusters with initial
densities of $>$10$^{2}$ M$_{\odot}$ pc$^{-3}$ preclude the production
of binaries with separations of $>$10$^4$ AU like Mizar-Alcor. Indeed,
Parker et al. (2009) conclude that ``[b]inaries with separations $>$
10$^4$ AU are 'always soft' - {\it any cluster will destroy such
binaries (if they could even form in the first place)}'' and that such
binaries must form in isolation. Mizar-Alcor would appear to be a
counter-example. Given the range of initial stellar densities probed
by the Parker et al. study, one can conclude that a reasonable upper
limit on the initial density of the UMa cluster is $<$10$^{2}$
M$_{\odot}$ pc$^{-3}$.

In comparing the Mizar-Alcor sextuplet to the known multiple star
population \citep{Tokovinin97, Eggleton08}, it appears that that
Mizar-Alcor (d $\simeq$ 25 pc) is the 2nd known closest multiple
system with 6 (or more) components after Castor (d $\simeq$ 16 pc).
The addition of Mizar-Alcor to the census of known multiple systems
with 6 or more components brings the census of such systems within 100
pc to 6, and effectively doubles the density of such systems within a
the 40 pc local volume.

\acknowledgements

We thank the Harvard-Smithsonian CfA TAC for allocating the MMT time
that made these observations possible.  We thank the MMT staff,
especially John McAfee, Alejandra Milone, Mike Alegria, and Tim
Pickering, and also Vidhya Vaitheeswaran and Thomas Stalcup for their
support of the MMT/AO system. We thank Eric Bubar for comments on the
manuscript. Clio is supported by grant NNG 04-GN39G from the NASA
Terrestrial Planet Finder Foundation Science Program. MAK is supported
by grant NNG 06-GE26G from the NASA Terrestrial Planet Finder
Foundation Science Program. EEM was supported by a Clay Postdoctoral
Fellowship at CfA during this observing program. MRM acknowledges
support through LAPLACE from the NASA Astrobiology Institute.


\end{document}